\documentclass[twocolumn,showpacs,preprintnumbers,amsmath,amssymb]{revtex4-1}

\usepackage{amsmath}

\usepackage{upgreek} 

\usepackage{epsfig, amsmath,amsfonts, amssymb,graphicx,amsthm,color}

\usepackage{mathtools}
\usepackage{dcolumn}
\usepackage{bm}
\usepackage{rotating}
\usepackage{amssymb}
\usepackage[latin1]{inputenc}
\usepackage{graphicx}
\usepackage{hyperref}

\usepackage{epstopdf}
\begin{document}

\title{Determining a vibrational distribution with a broadband optical source}
\author{T. Courageux, A. Cournol, D. Comparat, B. Viaris de Lesegno, H. Lignier}
\affiliation{Universit\'e Paris-Saclay, CNRS, Laboratoire Aim\'e Cotton, 91405, Orsay, France.}

\date{\today}

\begin{abstract}
This work presents an experimental protocol conceived to determine the vibrational distribution of barium monofluoride molecules seeded in a supersonic beam of argon. Here, as in many cases, the detection signal is related  to the number of molecules by an efficiency involving several parameters that may be difficult to determine properly. In particular, this efficiency depends on the vibrational level of the detected molecules. Our approach avoids these complications by comparing different detection signals generated by different vibrational excitations. Such an excitation is made possible by the use of a broadband optical source that depletes a specific vibrational level whose population is redistributed in the other levels. 
\end{abstract}

\maketitle
\section{Introduction}

In various experimental contexts, the understanding of physico-chemical mechanisms (shock wave dynamics \cite{Sutton1993}, plasma dynamics \cite{Staack2006}, laser ablation \cite{Yamagata1999}, buffer gas cooling \cite{Rellerget2013,Otto2013}) necessitates to characterize the evolution of the molecular degrees of freedom, which can be done by measuring the translational, rotational and vibrational temperatures. For example, in the field of molecular beams, it has been established for decades that a cooling process based on collisions is much more efficient for translation and rotation than for vibration \cite{mcclelland1979vibrational}. As a consequence, the vibrational temperature can be found much higher than the rotational and translational temperatures \cite{Wall2016}. 

There are various methods to measure the vibrational temperature but the possibility of using them may depend on the system under study. A frequently used technique, based on fluorescence spectroscopy, estimates the temperature through the relative intensity of the emission lines, assuming the vibrational populations follow the Maxwell-Boltzmann distribution \cite{Yamagata1999,Tarbutt2002a,Zhang2015}. Another possibility is to perform a photodissociation thermometry where  molecules lying in a given vibrational level are selectively dissociated \cite{Rellerget2013}.

In this work, we demonstrate that the vibrational distribution of barium monofluoride (BaF) molecules seeded in a supersonic beam of argon can be determined by combining a detection based on a resonant enhanced multiphoton ionization (REMPI) spectroscopy and a vibrational optical excitation. The determination of the vibrational distribution from the REMPI signal is not straightforward because the detection efficiency, denoted $e_{v_\mathrm{d}}$, should be calculated and varies with the detected vibrational level $v_\mathrm{d}$. To get rid of this difficulty, we used an optical excitation that is able to send all the molecules from a given vibrational level $v_\mathrm{e}$  to the nearby ones. This redistribution leads to modifications on the REMPI signal that a simple physical model can easily relate to the sought vibrational distribution.

The article is organized in three sections. The first one describes the main features of the experiment, in particular the broadband optical excitation and the REMPI detection. The second one presents the model used to calculate the vibrational distribution from the experimental data. Finally, in the last section, the experimental results are presented and discussed.

\section{Experiment}

\subsection{General description}
Our experiment is based on a 10 Hz pulsed supersonic beam produced by an adiabatic expansion of 5 bars of argon (with 2\% of SF$_6$) into a vacuum chamber ($4\times10^{-5}$ mbar) \cite{Cournol2018}. During the expansion, barium material is laser-ablated from a solid pellet and chemically reacts with SF$_6$ to produce BaF molecules. The BaF molecules are then cooled and entrained in the carrier gas beam which gives rise to a molecular beam. It penetrates, through a skimmer, into a second compartment of the vacuum chamber where the local pressure does not exceed $3\times10^{-7}$ mbar. After traveling a distance of 25 cm in about 440 $\mu$s, the BaF molecules are detected. 
During the propagation, the BaF molecules in the state $ \mathrm{X}^2 \Sigma_{1/2}^+$ can be submitted to a counter-propagating light beam emitted by a self-seeded tapered amplifier (TA). The interaction with this light excites the molecules to a selected vibrational level of the upper state $ \mathrm{A}^2 \Pi_{1/2}^+$ from which they relax to the X state. This process is the key element of our approach. Eventually, the BaF molecules lying the X state are selectively photoionized by an Optical Parametric Oscillator (OPO) pulse. An electrostatic device directs the ions to a stack of micro-channel plates whose signal is our measurement. The excitation process and the detection stage are detailed below.

\subsection{Broadband optical excitation}\label{Optical_excitation}

The TA source, developed in our laboratory, has been described in a previous publication \footnote{see Supplemental Material of Ref. \cite{Cournol2018}. Retrieved from \url{http://link.aps.org/supplemental/10.1103/PhysRevA.97.031401}}. Its central wavelength is adjustable between 845 nm ($\sim 11800$  cm$^{-1}$) and 860 nm ($\sim 11600$ cm$^{-1}$), its bandwidth is  $\sim 4$ cm$^{-1}$, and the optical power, dependent on the wavelength, ranges from 0.7 W to 1 W. As sketched in Fig. \ref{fig:sketch_spectra}, this light is used to excite the transitions $ \mathrm{X}^2 \Sigma_{1/2}^+(v_\mathrm{e}) \rightarrow \mathrm{A}^2 \Pi_{1/2}(v_\mathrm{A})$ where $v_\mathrm{e}$ and $v_\mathrm{A}$ are, respectively, the vibrational quantum numbers of the X and A states. All the transitions are achieved with $v_\mathrm{A}=v_\mathrm{e}-1$. The wavelengths associated with the different values of $v_\mathrm{e}$ are evenly spaced by about $28$ cm$^{-1}$ due to the fact that the X and A states are very similar \cite{Steimle2011,Chen2016}. The broad spectral width is necessary to excite all the molecules lying in a given $v_\mathrm{e}$ regardless of their rotational state, which is made possible by covering the Q rovibrational branch. 

The relaxation from the A state to the X state mainly results from a single photon process. Here we deliberately ignore the two photon process described in several studies \cite{Chen2016,Albrecht2020} as its effect is negligible. The redistribution of BaF in the vibrational levels of the X state follows the well-known Franck-Condon (FC) principle applied to the A-X transition. In its simplest form, the FC principle states that the transition electronic dipole moment has no variation with respect to the internuclear distance \cite{BernathChap9}. This approximation, supported by ab-initio calculations \cite{Kang2016,Yang2020}, implies that the part of the population decaying in a detected vibrational level $v_\mathrm{d}$ is essentially proportional to the overlap integral between the vibrational levels of the A and X states, the so-called Franck-Condon factors $q_{v_\mathrm{A},v_\mathrm{d}}$ (or equivalently $q_{v_\mathrm{e}-1,v_\mathrm{d}}$ in the context of the excitation as the relation between $v_\mathrm{A}$ and $v_\mathrm{e}$ is fixed). For BaF, the FC factors of the A-X transitions are particularly highly diagonal \cite{Chen2016,Hao2019a}, i.e., up to $v_\mathrm{e}=7
$, we have $0.4<q_{v_\mathrm{e}-1,v_\mathrm{d}}<0.95
$ for $(v_\mathrm{e}-1)-v_\mathrm{d}=0$, $0.04<q_{v_\mathrm{e}-1,v_\mathrm{d}}<0.28
$ for $|(v_\mathrm{e}-1)-v_\mathrm{d}|=1$ and  $q_{v_\mathrm{e}-1,v_\mathrm{d}}<0.06$ for $|(v_\mathrm{e}-1)- v_\mathrm{d}|>1$. 

From these considerations, it is expected that the de-excitation mainly increase the signal related to $v_\mathrm{d}=v_\mathrm{e}-1$ and some nearby levels. The only level for which the detection should drop is $v_\mathrm{d}=v_\mathrm{e}$ because it is emptied by a continuous TA excitation throughout the beam propagation. Indeed, even if molecules decay back to the initial vibrational level, the excitation time is long enough for them to be re-excited. As this return is not favored by the FC factors ($q_{v_\mathrm{e}-1,v_\mathrm{e}}< 0.28$ ), the depletion of the initial vibrational level only needs a few photons. 

We will show further (section \ref{Methodology}) how a careful analysis of the detection signal changes induced by the TA excitation allows us to reconstruct the vibrational distribution.

\subsection{Ionization spectroscopy}

\begin{figure*}[htp]
\includegraphics[width=0.45 \textwidth]{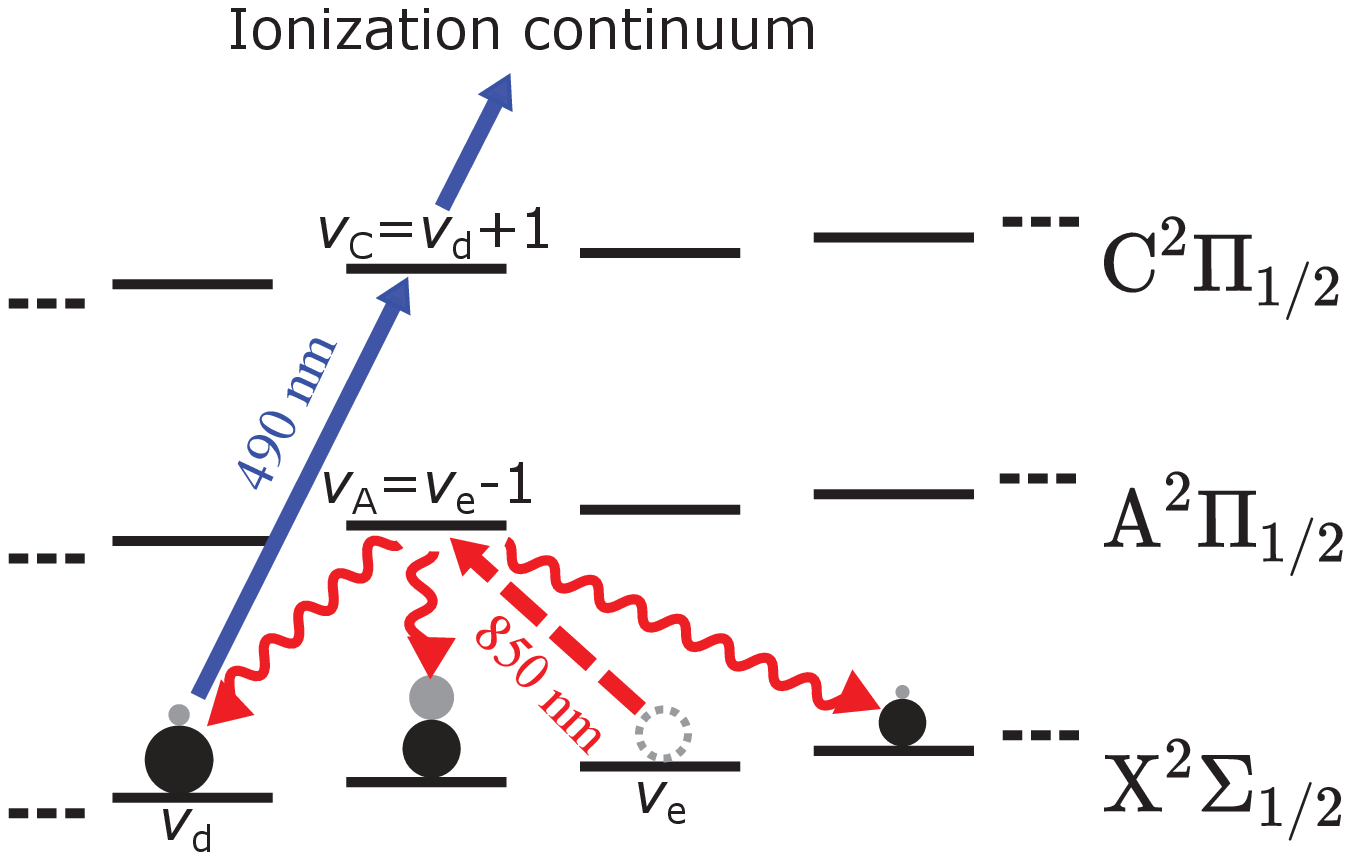} 	\hfill
\includegraphics[width=0.5 \textwidth]{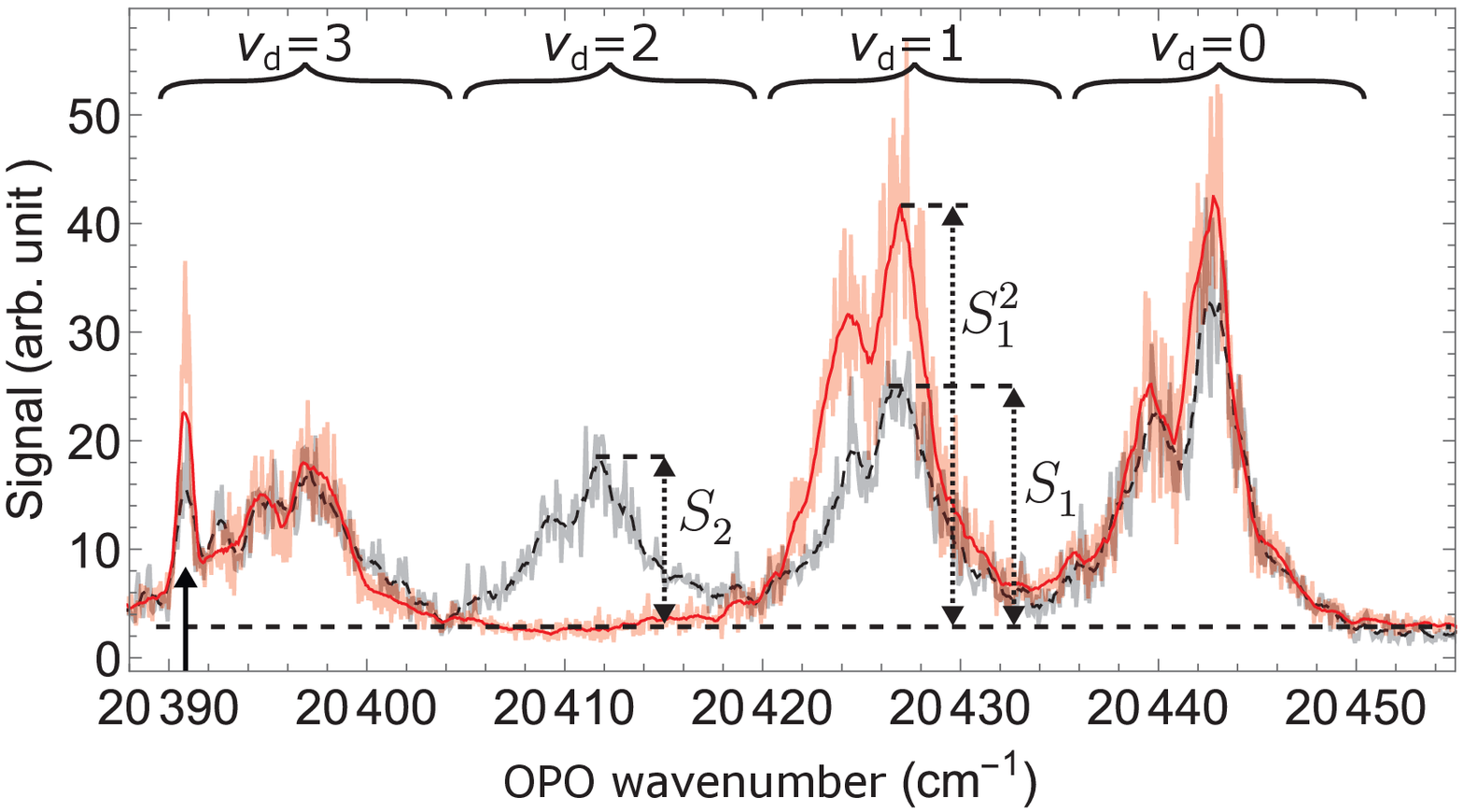}
   	\caption{Left: Schematics of the optical transitions used for the vibrational excitation and the detection by photo-ionization. The black horizontal segments represent a few successive vibrational levels of the electronic states indicated on the right hand side. A continuous broadband optical source (845-860 nm) excites the molecules (gray dashed circle) from $v_\mathrm{e}$  to $v_\mathrm{A}=v_\mathrm{e}-1$ (red dashed arrow). Due to spontaneous emission (wavy red arrows) from the A state, molecules are transferred to levels different from $v_\mathrm{e}$ (initial and incoming populations are respectively represented by dark and gray disks). The photo-ionization is a REMPI process induced by an OPO pulse using resonant transitions between $v_\mathrm{d}$ and $v_{\mathrm{C}}=v_\mathrm{d}+1$ (blue plain upwards arrow). The transition to the ionization continuum is not shown. Right: Two photo-ionization spectra recorded with an OPO pulse energy of 5 mJ showing the four first bands $v_\mathrm{d}=0,1,2,3$. The raw data points are displayed along with a moving average over 10 points in case of an excitation of the level $v_{\mathrm{e}}=2$ (black, dashed line) and in absence of excitation (red, plain line). The band $v_{\mathrm{d}}=2$ is depleted when the excitation light is on. The P and Q rovibrational branches and two associated cusps are visible in each band (the line indicated by the black arrow is a parasitic atomic line). $S_2$, $S_1^2$ and $S_1$ are illustrative examples of the quantities used in the main text.}  
   		\label{fig:sketch_spectra}	
\end{figure*}

 The REMPI detection relies on the absorption of two photons from the OPO. The first one resonantly excites the transition $ \mathrm{X}^2 \Sigma_{1/2}^+(v_\mathrm{d}) \rightarrow \mathrm{C}^2 \Pi_{1/2}(v_\mathrm{C})$ where $v_\mathrm{d}$ is the detected vibrational level of the X state and  $v_\mathrm{C}$ is the vibrational quantum number of the C state. This optical transition is sketched in Fig. \ref{fig:sketch_spectra}. The second photon drives the molecules from the C state to the ionization continuum. The OPO pulse has a typical width of 0.1 cm$^{-1}$ and its wavelength can be adjusted between 489 nm (20450 cm$^{-1}$) and 492 nm (20325 cm$^{-1}$). 
 
Our photo-ionization spectra are measured by scanning the wavelength of the OPO in the range covering the transitions such that $v_\mathrm{C}=v_\mathrm{d}+1$. Figure \ref{fig:sketch_spectra} shows two such spectra covering the first four vibrational levels ($v_\mathrm{d}=0,1,2,3$). The first spectrum was obtained in absence of vibrational excitation while the second was taken with a broadband excitation of $v_\mathrm{e}=2$. The excitation causing a complete depletion, we observe a disappearance of the corresponding band. On the  other hand, the signal related to the bands $v_\mathrm{d} = 1$ and, to a lesser extent $v_\mathrm{d}=0$, have increased. 

A photo-ionization spectrum is essentially structured by the
resonant transitions between the X and C states. The rovibrational lines are not resolved because of the laser linewidth and the even stronger power broadening effects. Instead, we observe broad peaks that correspond to rovibrational branches. The most intense peak is related to the Q-branch, spread over nearly 2 cm$^{-1}$ (henceforth called Q peak) while the second peak, slightly broader and less intense, corresponds to the P branch. The two peaks partially overlap but we can see in Fig. \ref{fig:sketch_spectra} that two cusps arise from this structure. 

From the distance between the cusps, we can easily extract the rotational temperature, i.e. how the molecules are distributed over the rotational levels, thanks to a comparison with a theoretical spectrum calculated by the software PGOPHER \cite{pgopher2017}. For all our data, we found that $T_\mathrm{rot}\approx 30$ K with less than 10\% of uncertainty (provided the parameters of ablation are kept unchanged). Note that this method requires to record a few hundred points to obtain a spectrum, which assumes that the rotational distribution does not fluctuate significantly over the experimental realizations.

\begin{figure}

\includegraphics[width=0.9 \columnwidth]{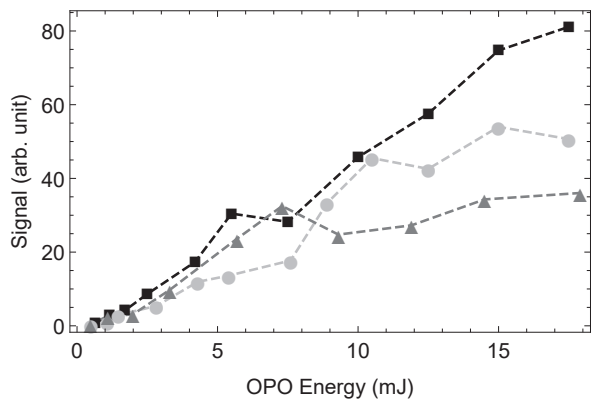}
   	\caption{Q-peak maximums relative to $v_\mathrm{d}=0$ (squares), $v_\mathrm{d}=1$ (triangles) and $v_\mathrm{d}=2$ (circle) as a function of the OPO energy used for photo-ionization. The vibrational distribution of molecules is approximately the same for each measurement. All the signals increase with the OPO energy but differently according to the detected level $v_\mathrm{d}$. This shows that the vibrational distribution cannot be simply inferred from the Q-peak maximums. The dashed lines are guidelines for the eye.}
   	
   	\label{fig:SignalvsPulseEnergy}
   	
\end{figure}

Paradoxically, although the vibrational bands are well separated and the rovibrational lines are unresolved, the determination of the population distribution among the vibrational levels is more difficult than the determination of the rotational temperature. This is due to the fact that the ionization strength varies with respect to the vibrational number and furthermore affects the power broadening effects: the peak maximums or the peak areas cannot be simply compared with each others. We illustrate this situation in Fig. \ref{fig:SignalvsPulseEnergy} where the Q peak maximums of $v_{\mathrm{X}}=0,1,2$ are plotted with respect to the OPO pulse energy. Any possible signal ratio between the Q peak maximums depends on the OPO energy and, as a consequence, cannot be equivalent to a ratio of vibrational populations, that should be constant.

\section{Methodology}\label{Methodology}

\subsection{Relationship between signals and vibrational populations}

We now describe the physical model that can be used to deduce the vibrational populations in $v_\mathrm{d}$, denoted $P_{v_\mathrm{d}}$ from the detection signals potentially modified by optical excitations described in \ref{Optical_excitation}.

Without excitation, we suppose to have access to a detection signal $S_{v_\mathrm{d}}$ proportional to the population $P_{v_\mathrm{d}}$. However the coefficient of proportionality $e_{v_\mathrm{d}}$ is unknown and depends on the level $v_\mathrm{d}$ and the detection parameters. The redistribution caused by the optical excitation is the key process allowing us to compensate for this lack of information.

When the excitation is enabled, we add the reasonable supposition that the relation of proportionality between signal and population holds with the same coefficient $e_{v_\mathrm{d}}$, which is written
\begin{equation}\label{eq:SignalvsPop} 
S_{v_\mathrm{d}}^{v_\mathrm{e}}=e_{v_\mathrm{d}} P_{v_\mathrm{d}}^{v_\mathrm{e}}
\end{equation}
where $S_{v_\mathrm{d}}^{v_\mathrm{e}}$ and $P_{v_\mathrm{d}}^{v_\mathrm{e}}$ are the signal and the population related to the level $v_\mathrm{d}$ while the optical excitation has been carried out on the level $v_\mathrm{e}$.

By using the FC principle, we then establish the relation between $P_{v_\mathrm{d}}^{v_\mathrm{e}}$ and the unexcited populations: 
\begin{equation} \label{eq:PopRelation}
P_{v_\mathrm{d}}^{v_\mathrm{e}} = 
\begin{cases}
P_{v_\mathrm{d}}+\alpha_{v_\mathrm{d},v_\mathrm{e}}P_{v_\mathrm{e}} & \mathrm{\ for\ }v_\mathrm{d} \neq v_\mathrm{e}  \\
0 & \mathrm{\ for\ } v_\mathrm{d} = v_\mathrm{e}
\end{cases}
\end{equation}
where 
\begin{equation}\label{eq:PopFraction}
\alpha_{v_\mathrm{d},v_\mathrm{e}}=\frac{q_{v_\mathrm{e}-1,v_\mathrm{d}}}{1-q_{v_\mathrm{e}-1,v_\mathrm{e}}}
\end{equation}
is the fraction of $P_{v_\mathrm{e}}$ ending up in the level $v_\mathrm{d}$ according to a simple rate equation model.

Finally, a manipulation of the expressions (\ref{eq:SignalvsPop})-(\ref{eq:PopFraction}) with the sum rule $\sum_{v_\mathrm{d}} q_{v_\mathrm{e}-1,v_\mathrm{d}}=1$ provides the system of algebraic equations
\begin{equation}\label{eq:Pop_with_matrix}
\sum_{v_\mathrm{d}} \left( \frac{S_{v_\mathrm{d}}^{v_\mathrm{e}}}{S_{v_\mathrm{d}}}-1\right)P_{v_\mathrm{d}}=0
\end{equation}
that can be put in matrix form. The kernel, numerically computed, is a solution vector whose components $P_{v_\mathrm{d}}$ are proportional to the vibrational populations. In the following, our results will always be presented as $P_{v_\mathrm{d}}/P_{0}$, i.e. the populations are expressed relatively to the population in the vibrational ground state.

\subsection{Determination of the experimental observable}

In order to extract the vibrational populations from Eq. (\ref{eq:Pop_with_matrix}), it is important to determine which kind of measurement can satisfyingly represent $S_{v_\mathrm{d}}$ and $S_{v_\mathrm{d}}^{v_\mathrm{e}}$. We must ensure that the chosen experimental quantities respect the conditions $e_{v_\mathrm{d}}=S_{v_\mathrm{d}}/P_{v_\mathrm{d}}=S_{v_\mathrm{d}}^{v_\mathrm{e}}/P_{v_\mathrm{d}}^{v_\mathrm{e}}$ established previously. We will see that the Q-peak maximum fulfills these requirements. 

The Q-peak maximum does not result from the ionization of all the molecules lying in $v_\mathrm{d}$. In fact, only a few rotational states, selected by the OPO paramaters (wavelength, fluence), contribute to the ion signal. However, if the rotational temperature is stable and the OPO parameters are fixed, the coefficient $e_{v_\mathrm{d}}$ is well determined and constant.

This description seems satisfying in absence of optical excitation but the transfer of population from $v_\mathrm{e}$ to $v_\mathrm{d}$ could undermine the validity of our approach. In fact, if the rotational distribution of the transferred population is too different from that of the population initially in $v_\mathrm{d}$, $e_{v_\mathrm{d}}$ will depend on the optical excitation. Therefore, our approach requires that the rotational distribution be almost independent of the vibration level and barely modified by the absorbed and emitted photons. Experimentally, it can be quickly verified that the optical excitation mainly affects the amplitude and not the shape of the REMPI signal for $v_\mathrm{d}$ (see Fig. \ref{fig:sketch_spectra}). This is consistent with simple considerations:  given that a single photon can change the molecular rotation by a quantum of angular momentum at most, the impact of the excitation can be evaluated through the number of photons necessary to empty $v_\mathrm{e}$. We saw previously that this number is small and, therefore, cannot substantially modify the rotational distribution spread over about 20 rotational levels. We can thus conclude that the Q-peak maximum is an acceptable quantity for our analysis developed above.

\section{Results and discussion}

\begin{figure}

\includegraphics[width=.9 \columnwidth]{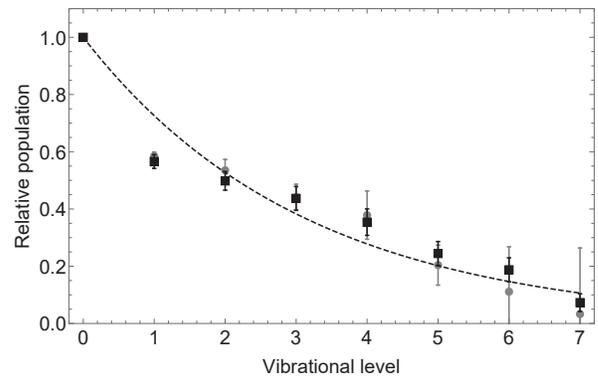}
   	\caption{Vibrational distributions of BaF seeded in a supersonic beam of argon. Two different methods are compared. The first is based on the resolution of Eq. \ref{eq:Pop_with_matrix} using experimental data only (squares). The second method, based on Eq. \ref{eq:Pop_with_FC}, combines the experimental data and theoretical values of FC factors (circles). Fitting these curves by Maxwell Boltzmann distribution (dashed line curve) gives a temperature of $2100 \pm 500$ K.}
   	
   	\label{fig:vibrational_distribution}
   	
\end{figure}

We collected several hundred independent realizations $S_{v_\mathrm{d}}^{v_\mathrm{e}}$ and $S_{v_\mathrm{d}}$ under conditions of BaF formation that were stabilized as much as possible. This operation took 2 hours approximately. All the data were organized in 7 series associated with the excitation of level $1\leq v_{\mathrm{e}}\leq 7$.  We systematically probed $v_{\mathrm{d}}$ for $v_{\mathrm{e}}-2\leq v_{\mathrm{d}}\leq v_{\mathrm{e}+1}$ with and without excitation (except for $v_{\mathrm{e}}=1$ where $v_{\mathrm{d}}=0,1,2$ and $v_{\mathrm{e}}=7$ where $v_{\mathrm{d}}=8$ was not retained due to truncation). For the other values of $v_{\mathrm{d}}$, we only quickly verified that the signal was not visibly affected by the excitation. This was anticipated because the proportion of excited molecules decaying in any of these levels is small, as is the related FC factor (see section \ref{Optical_excitation}). We also took care to check that the excited level was almost completely depleted (clearly visible in Fig. \ref{fig:sketch_spectra}), as required by our data analysis (see Eq. (\ref{eq:PopRelation})).
The raw data were corrected to account for the progressive decrease of the total number of molecules with respect to the number of ablation shots. To this end, we regularly acquired $S_0$ a reference measurement. Its time evolution was properly fitted by a function used to re-scale our data, an operation leading to a correction smaller than 10\%. We also used this reference measurement to normalize the signals of interest.

To determine the experimental populations, we used the system of equations (\ref{eq:Pop_with_matrix}) but $S_{v_\mathrm{d}}^{v_\mathrm{e}}$ and $S_{v_\mathrm{d}}$ were replaced by the experimental mean values $\bar{S}_{v_\mathrm{d}}^{v_\mathrm{e}}$ and $\bar{S}_{v_\mathrm{d}}$ obtained by a systematic averaging of at least 400 points. Choosing the mean values rather than the individual measurements allowed us to avoid complications caused by the fact that all the points were measured independently and subject to strong fluctuations  (in the order of 30\%) that are assumed to be caused by variations on the OPO energy and the number of formed molecules. We conducted Monte-Carlo simulations to verify that this approach was able to provide the correct vibrational population distributions. Besides, the population uncertainties were estimated by standard error propagation methods as explained, for example, in \cite{Tanabashi2018}. 

From $\bar{S}_{v_\mathrm{d}}^{v_\mathrm{e}}$ and $\bar{S}_{v_\mathrm{d}}$, we calculated the terms in parenthesis of Eq. (\ref{eq:Pop_with_matrix}). The terms corresponding to $v_\mathrm{d}=v_\mathrm{e}$ were equal to -1 as a consequence of the full depletion ($\bar{S}_{v_\mathrm{e}}^{v_\mathrm{e}}=0$). The vast majority of the other terms were found positive which was expected because any population in $v_\mathrm{d}\neq v_\mathrm{e}$ should only increase ($\bar{S}_{v_\mathrm{d}}^{v_\mathrm{e}}>\bar{S}_{v_\mathrm{d}}$). However, we found that some terms, especially those associated with $v_{\mathrm{d}}=v_{\mathrm{e}}+1$ were slightly but frequently negative and corresponded to a population decrease smaller than 10\%. A probable explanation is that the broadband excitation, that should have been limited to the band $v_{\mathrm{e}}$, partially reached the level $v_{\mathrm{e}}+1$. Despite this somewhat puzzling data, we determined the vibrational distribution, solution of the Eq. (\ref{eq:Pop_with_matrix}), which is shown in Fig. \ref{fig:vibrational_distribution}.

\begin{figure}

\includegraphics[width=1.0 \columnwidth]{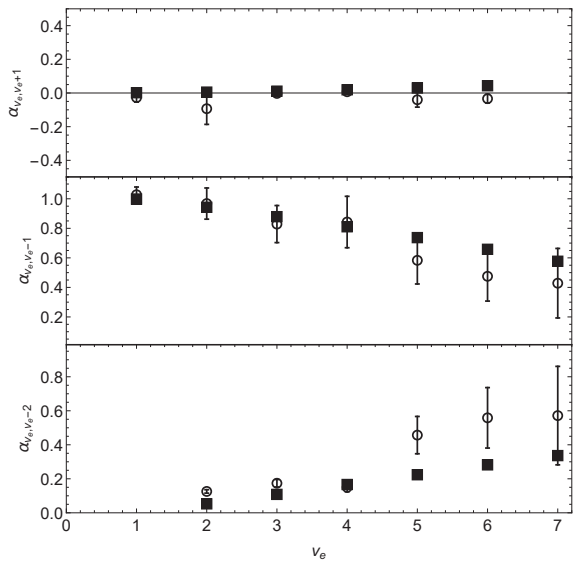}
   	\caption{Coefficients  $\alpha_{v_\mathrm{e},v_\mathrm{e}+1}$ (top), $\alpha_{v_\mathrm{e},v_\mathrm{e}-1}$ (middle) and $\alpha_{v_\mathrm{e},v_\mathrm{e}-2}$ (bottom) as a function of the vibrational level  $1\leq v_\mathrm{e}\leq7$. The values calculated from the theoretical FC factors (squares) are compared to the experimentally ones (circles).}
   	 	\label{fig:coefficients}
\end{figure}

To test the robustness of our method, our distribution can be compared to that obtained with another analysis that does not use the terms related to $v_\mathrm{e}=v_\mathrm{e}+1$ but requires an a priori knowledge of the FC factors of the A-X transition. Because the FC factors in the literature are not given for sufficiently high vibrational quantum numbers, we calculated them with a Morse Potential Model applied to the experimental data published in \cite{Effantin1990a}. With the additional information provided by this approach, the system of Eq.(\ref{eq:Pop_with_matrix}) is over-determined. It is then possible to make an analysis using a restricted set of experimental data, namely $\bar{S}_{v_\mathrm{d}}^{v_\mathrm{e}}$ and $\bar{S}_{v_\mathrm{d}}$ where $v_\mathrm{d}=v_\mathrm{e}-1$ only. The particular choice of $v_\mathrm{d}$ allowed us to minimize the uncertainties on the populations because this levels is subject to the maximum transfer of population (see section \ref{Optical_excitation}). 

Mathematically, the determination of the vibrational distribution relies on a simple manipulation of Eq. (\ref{eq:PopRelation}) that gives the population ratio between adjacent levels
\begin{equation}\label{eq:Pop_with_FC}
\frac{P_{v_\mathrm{e}}^\mathrm{exp}}{P_{v_\mathrm{e}-1}^\mathrm{exp}}=\frac{1}{\alpha^{\mathrm{th}}_{v_\mathrm{e}-1,v_\mathrm{e}}}\left(\frac{\bar{S}^{v_\mathrm{e}}_{v_\mathrm{e}-1}}{\bar{S}_{v_\mathrm{e}-1}}-1\right).
\end{equation}
where the coefficient $\alpha^{\mathrm{th}}_{v_\mathrm{e}-1,v_\mathrm{e}}$ are the terms (\ref{eq:PopFraction}) evaluated with the theoretical FC factors. As this population ratio can be obtained for all the values taken by $v_\mathrm{e}$, we have a recurrence relation from which we extracted the population distribution displayed in Fig. \ref{fig:vibrational_distribution}. Except for the populations in $v_{\mathrm{X}}=6,7$, the two vibrational distributions turn out to be in very good agreement.

To go further, we also compared the coefficients $\alpha^{\mathrm{th}}_{v_\mathrm{d},v_\mathrm{e}}$ and the experimental ones deduced by
\begin{equation}
\alpha_{v_\mathrm{d},v_\mathrm{e}}^\mathrm{exp}=\frac{P_{v_\mathrm{d}}^\mathrm{exp}}{P_{v_\mathrm{e}}^\mathrm{exp}}\left(\frac{\bar{S}_{v_\mathrm{d}}^{v_\mathrm{e}}}{\bar{S}_{v_\mathrm{d}}}-1\right)
\end{equation}
The comparison in Fig. \ref{fig:coefficients} shows that the experimental and theoretical values are in rather good agreement, especially for low vibrational levels. The discrepancy especially affects the higher levels, which can be at least partially explained by an effect of error accumulation inherent to our method. 
However, the overall good adequacy tends to justify our approximations and hypotheses and consolidates the validity of the vibrational distribution determined previously. 

We finally fitted the vibrational distribution with a Maxwell-Boltzmann distribution, also plotted in Fig. (\ref{fig:vibrational_distribution}) for comparison. This gave us a vibrational temperature $T_\mathrm{vib}=2100\pm 500$ K that must be considered as a rough quantity. However, it is much larger than the rotational temperature $T_\mathrm{rot}\approx 30$ K previously mentioned. Interestingly, the ratio between these internal temperatures is close to that found for Ytterbium monofluoride (YbF) seeded in a supersonic beam of argon \citep{Tarbutt2002a}, albeit the YbF internal temperatures were 10 times lower. We find here the effect already reported that the vibration is not thermalized as well as the rotation. This incomplete thermalization might explain why the experimental distribution slightly deviates from the Maxwell-Boltzmann distribution.

\section{Conclusion}

In conclusion, we demonstrated that a broadband optical source is a relevant tool to determine the relative vibrational populations of a sample of molecules. Unlike other techniques relying only on fluorescence or ionization spectrum analysis, our approach does not require to calculate the detection efficiency as it is limited to a comparison of the detection signals. It should also be noted that the REMPI detection used in this work is not inherent to our approach; for example, it should be possible to use a fluorescence-based technique. Finally, we think our method is applicable to diatomic molecules other than BaF but an additional experimental effort is to be envisaged if the FC factors are not highly diagonal. In fact, if all the FC factors turn out to be much smaller than one, a large number of vibrational levels must be detected. In this case, the changes induced by the excitation are expected to be weak compared than what we observed in this work, which should imply an averaging over a large amount of data. Also, it is probable that complex excitation spectra whose shape could depend on $v_\mathrm{e}$ might be required. 

\section{Conflicts of interest}

There are no conflicts to declare.

\section{Acknowledgments}

The research leading to these results has received funding from ANR MolSysCool, Dim Nano K CPMV.

\bibliographystyle{h-physrev}

\bibliography{references}

\end{document}